\documentclass[
  showpacs,
  twoside,
  twocolumn,
  aps,
  prlsds
]{revtex4}

\usepackage[ngerman,american]{babel}
\usepackage{graphicx}
\usepackage{amsmath}
\usepackage{pxfonts}
\usepackage{wasysym}
\usepackage{manfnt}
\usepackage{color}
\usepackage{dingbat}
\usepackage{multirow}
\begin{document}

\title{Percolation of disordered jammed sphere packings}

\author{Robert M. Ziff}
\email[email: ]{rziff@umich.edu}
\affiliation{Center for the Study of Complex Systems and Department of Chemical Engineering, University of Michigan, Ann Arbor, Michigan 48109-2136, USA}
\author{Salvatore Torquato}
\email[email: ]{torquato@princeton.edu}
\affiliation{Department of Chemistry, Department of Physics, Princeton Institute for the Science and Technology of Materials, and Program in Applied and Computational Mathematics, Princeton University, Princeton, New Jersey 08544, USA
}

\begin{abstract}
We determine the site and bond percolation thresholds for a system of disordered jammed sphere packings in the maximally random jammed state, generated by the Torquato-Jiao algorithm.    For the site threshold, which gives the fraction of conducting vs.\ non-conducting spheres necessary for percolation, we find $p_c = 0.3116(3)$, consistent with the 1979 value of Powell $0.310(5)$
and identical within errors to the threshold for the simple-cubic lattice, 0.311608, which shares the same average coordination number of 6. In terms of the volume fraction $\phi$, the threshold corresponds to a critical value $\phi_c = 0.199$.   For the bond threshold, which  apparently was not measured before, we find $p_c = 0.2424(3)$.  To find these thresholds, we considered two shape-dependent universal ratios involving the size of the largest cluster, fluctuations in that size, and the second moment of the size distribution; we confirmed the ratios' universality by also studying the simple-cubic lattice with a similar cubic boundary.  The results are applicable to many problems including conductivity in random mixtures, glass formation, and drug loading in pharmaceutical tablets.
\end{abstract}



\maketitle

\section{Introduction}
\label{sec:introduction}
When spheres are packed in a disordered jammed arrangement, they are in direct contact with each other at several points, and create a contact network.  The number of neighbors of each node in this network is between four and twelve with an average of 6, not counting non-jammed spheres---the so-called rattlers \cite{TorquatoStillinger10}.  A natural question to ask is what the percolation threshold of this network is \cite{StaufferAharony94}.  That is, say a certain fraction $p$ of the spheres are conducting and the rest are non-conducting.  What is the value of $p$ at which long-range conduction first occurs?  For the contact network or graph, this quantity corresponds to the site percolation threshold.  One could also ask for the bond threshold of the network.  These are fundamental and practically important questions about packings, which however have barely been addressed numerically.

Practical applications of knowing the threshold for packed particles include formation of glasses in binary metallic systems \cite{TakeuchiInoue12}, combustion in packed mixtures \cite{FrolovPivkinaAleshin97}, and controlling conductivity in a  electric arc furnace \cite{ChuNiuLiuWang03}.  
An important application is in the design of pharmaceutical tablets, where the composition can be a packed mixture of soluble and insoluble particles.  Percolation has been used extensively in the design of drug release and tablet compaction \cite{SiegelKostLanger89,LeuenbergerHolmanUsteriWinzap89,Leuenberger99,LeuenbergerBonnyKolb95,SorianoCaraballoMillanPineroMelgozaRabasco98,MillanCaraballo06,KimuraPuchkovBetzLeuenberger07}.  This work has relied substantially on the percolation properties of simple lattices.

From a theoretical point of view, it is useful to find the thresholds of various kinds of networks.  While thresholds on numerous three-dimensional lattices have been extensively studied (e.g., \cite{LorenzZiff98,AcharyyaStauffer98,BallesterosEtAl99,TranYooStahlheberSmall13,XuWangLvDeng14,WangZhouZhangGaroniDeng13,Malarz15}), most have been for regular lattices of constant coordination number.   The disordered jammed network provides an example of a three-dimensional system with a wide distribution of nearest neighbors, complementing the Poisson-Voronoi tessellation network \cite{JerauldScrivenDavis84} and the Penrose quasi-crystal \cite{ZakalyukinChizhikov05,Henley86}, whose percolation thresholds have both been studied.  Other recent works on percolation thresholds include Refs.\ \cite{BormanGrekhovTroninTronin15,Jacobsen14,Jacobsen15,DamavandiZiff15,HajiAkbariHajiAkbariZiff15,ScullardJacobsen16}.

The first determinations of the site threshold of the disordered jammed hard-sphere network were done experimentally---conducting and insulating balls of various materials were randomly packed and the critical fraction $p_c$ for long-range conductivity to occur was measured.  In 1974,  Fitzpatrick, Malt and Spaepen \cite{FitzpatrickMaltSpaepen74} measured $p_c = 0.27$ for a mixture of 5000 aluminum and acrylic spheres.   In 1978, Ottavi et al.\ \cite{OttaviClercGiraudRoussenqGuyonMitescu78} found $p_c \approx 0.30$ by studying a system of $\approx10^5$ conducting and insulating molded ABS plastic spheres.  This was followed in 1979 by the first Monte-Carlo study of connectivity of a disordered jammed packed-sphere system, carried out by 
Powell \cite{Powell79,Powell80}, who determined the site thresholds for packings created by the method of Tory et al.\ \cite{ToryEtAl73}  and Matheson \cite{Matheson74} where spheres are deposited vertically in a gravitation-like field.  Powell found $p_c = 0.310(5)$, where the number in parenthesis is the error in the last digit, and noticed that this was within error bars of the threshold $\approx0.3116$ for site percolation on the cubic lattice, which has a uniform coordination number of 6.
In 1982,  a hybrid experimental/simulation study was performed by 
Ahmadzadeh and Simpson \cite{AhmadzadehSimpson82}, who carried out simulations using the experimentally derived contact network of Bernal, and found $p_c = 0.32$.  In 1986, Oger et al.\ \cite{OgerTroadecBideauDoddsPowell86} experimentally measured a threshold of 0.29 using a system of glass beads, some silver-coated, which deformed somewhat by the pressure of the packing.   In 1996, Sunde found a threshold of about 0.30 by a Monte-Carlo simulation of packing around a seed of contacting spheres \cite{Sunde96}.   There has been more recent work studying percolation of mixtures of sphere of different radii \cite{BerteiNicolella11}.  As can be seen, most of this work is rather old and of relatively low precision, and only concerns site percolation; as far as we can tell no measurements have been made of the bond threshold of the jammed sphere contact network. 

The goal of this paper is to find the site and bond thresholds of the disordered jammed packed-sphere network to relatively high precision, and to explore efficient numerical methods to find precise thresholds for such systems for future work.

A complicating factor in this endeavor is that in simulations, as in real systems, randomly packed spheres can have different final packing fractions and correlation properties depending upon the method of packing \cite{TorquatoStillinger10}.  Torquato, Truskett and Debenedetti \cite{TorquatoTruskettDebenedetti00} introduced a mathematically precise definition of a maximally random jammed (MRJ) state as the most disordered packing subject to the condition that it is jammed (mechanically stable).  Torquato and Jiao (TJ) \cite{TorquatoJiao10}  devised an algorithm that produces dense random packings close to this state.  Other protocols can be devised that have different packings \cite{JiaoStillingerTorquato11}.  A natural question to ask is how the different packings affect the percolation properties.  As a first step in this program, we study the precise site and bond thresholds of packings produced by the TJ algorithm, which gives packing of density of 0.639 \cite{AtkinsonStillingerTorquato13}.

An important early development in the understanding of percolation thresholds was the introduction of the critical volume fraction (CVF or  $\phi_c$)  by Scher and Zallen \cite{ScherZallen70}.  For regular lattices with uniform nearest-neighbor bond lengths of unity, they considered placing spheres of unit diameter at each vertex; the fraction of space occupied by each sphere gives the filling fraction $f$.  Multiplying this quantity by the site percolation threshold $p_c$ yields $\phi_c = p_c f$.  For the three-dimensional systems that were studied, it was found the $\phi_c$ falls in a narrow range of 0.144-0.163.  Likewise, one can define a critical filling fraction for the disordered jammed spheres as the fraction of space occupied by the conducting spheres at the critical point.  Using the packing density $f = 0.59$, Powell found $\phi_c =  0.183$, somewhat higher than values for other three-dimensional systems.   Finding $\phi_c$ for 
the MRJ system is another motivation for this work.  

We mention that percolation of jammed particles on discrete lattices has also been studied, for example, the percolation of dimers in two and three \cite{TarasevichCherkasova07} dimensions, and percolation of rectangles adsorbed by random sequential adsorption \cite{KriuchevskyiBulavinTarasevichLebovka14}.  In the continuum, it seems that packed spheres is the only jammed system whose percolation behavior has been studied.

In the following sections we discuss invariant ratios that we use, describe the simulation methods, give the results of the study on the MRJ, and for comparison present results for a simple cubic lattice.  We also investigate a two-dimensional system to make contact with some previous work.  We close with a discussion of the results.

\section{Methods}

To carry out this work, we analyzed three unique samples of $N = 2000,$ 6000, and 10000 MRJ sphere packings, generated by the TJ algorithm.  The overall system shapes were cubic with periodic boundary conditions in all directions.  
In a random packing, a small fraction of the spheres constitute the rattlers that are not locked into the structure.
For the three random packings considered here, the number of rattlers were 34, 87,  and 130 respectively---about 1.5 \% of the total number of spheres.  We assume that these rattlers do not contribute to the percolating networks, even though a conducting rattler could sporatically connect two paths.  On the other hand, we include the rattlers when we consider the filling factor $f$ of the networks, though including the rattlers does not change $p$ (the fraction of conducting spheres) since the rattlers are conducting or insulating spheres with the same probability as the rest of the spheres.

To measure the percolation threshold for such relatively small systems is somewhat of a challenge.  One possible approach would be to find the probability of the existence of a wraparound cluster, which is known to lead to rapidly converging estimates for $p_c$ \cite{NewmanZiff00,NewmanZiff01}.  However, the data files that we used only contained neighbor lists of each indexed particle with no information of position or where the periodic boundaries were crossed.  (This sort of neighbor-list file is common for networks.)  Consequently, we could not use the wraparound method here.

A second approach would be to look at the  distribution $n_s(p)$ of clusters of size $s$ at different values of $p$ and find the value of $p$ such that $n_s \sim s^{-\tau}$ where $\tau \approx 2.1891$ for three-dimensional percolation 
\cite{BallesterosEtAl99,XuWangLvDeng14}.  We attempted this here; however, for such small systems, the finite-size effects overwhelm the power-law behavior and makes this method inaccurate.

A third method, which we follow here, is to study dimensionless quantities related to the cluster statistics that are universal at the critical point, similar to the Binder ratios used in studies of critical phenomena.  One property that we consider is the ratio of the square of the size of the largest cluster to the second moment
\begin{equation}
R_1 = \frac{N P_\infty^2}{S} = \frac{\langle s_\mathrm{max}\rangle^2}{\sum_i s_i^2}
\end{equation}
where $P_\infty =\langle s_\mathrm{max} \rangle / N$, $s_i$ is the number of sites in the $i$-th cluster,  and the second moment $S$ is
\begin{equation}
S = \frac1N \sum_{i=1}^\infty s_i^2
\end{equation}
A second quantity we consider is the ratio of the fluctuations in the largest cluster size to the square of the largest cluster size:
\begin{equation}
R_2 = \frac{\langle s_\mathrm{max}^2 \rangle - \langle s_\mathrm{max} \rangle ^2}{\langle s_\mathrm{max}\rangle^2}
\end{equation}
The universality of $R_1$ was discussed by Aharony and Stauffer \cite{AharonyStauffer97} and follows from the finite-size scaling of $P_\infty \sim L^{-\beta/\nu}$, $S \sim L^{\gamma/\nu}$, $N = L^{d}$ with the critical exponents satisfying the hyperscaling relation $2 \beta + \gamma = d \nu$ (for $d \le 6$)   where $d$ is the dimensionality.  The universality of $R_1$ can also be seen from the fact that it is the ratio of the square of largest cluster divided by a sum that is dominated by the squares of large clusters.  The large clusters will be similar  for different  systems of the same shape at the critical point, apart from system-dependent metric factors that cancel out, so this ratio is  {\it prima facie}  universal.  Likewise $R_2$ also depends on ratios of squares of largest clusters and is also expected to be universal.  However, universality of these two quantities is limited in the sense that their values depend upon the shape of the boundary of the system, and the boundary conditions, since those things will affect the properties of the large clusters.  This shape-dependent universality is seen in other quantities, such as a ratio related to properties of clusters connected to the origin in a finite system \cite{deSouzaTomeZiff11}, crossing probabilities \cite{LanglandsPichetPouliotSaintAubin92}, and others \cite{AharonyStauffer97}.   Thus, to compare values of $R_1$ and $R_2$ to other percolating systems, it is necessary to consider systems of the same overall shape and boundary condition. 

Fluctuations in the largest cluster near $p_c$ were previously discussed by Coniglio and Stauffer \cite{ConiglioStauffer80,ConiglioStanleyStauffer79,Stauffer80}, who showed that $\langle s_\mathrm{max}^2 \rangle - \langle s_\mathrm{max} \rangle ^2$ scales as $|p-p_c|^\gamma$, the same as $ \langle s_\mathrm{max} \rangle ^2$ and the susceptibility $\chi$ near the critical point, implying that $R_2$ goes to a finite value.   
They did not discuss the  behavior for $p$ away from $p_c$, nor the shape-dependent universality of this quantity.  As a test of our program, we also consider this 2D system below and find similar results.

To carry out the measurement, we use the algorithm of Newman and Ziff (NZ) \cite{NewmanZiff00,NewmanZiff01}, in which sites or bonds are made occupied one at a time in random order, and a record is kept of various quantities ($s_\mathrm{max}$, $s_\mathrm{max}^2$, $S$) as a function of the number of occupied sites or bonds $n$.  This is the so-called microcanonical value.  Then, to find the canonical value representing a system for a given $p$, we convolve with a binomial distribution, though for larger systems the difference between the microcanonical and canonical values can be quite small.  For a microcanonical quantity $Q_n$, the convolution to the canonical $Q(p)$ is given by  
\begin{equation}
Q(p) = \sum_{n=0}^N {N \choose n} p^n (1-p)^{N-n} Q_n
\end{equation}
Methods to carry out the convolution, including how to calculate the binomial distribution by using recursion, are described in \cite{NewmanZiff01}.  The results of the microcanonical calculations are stored in one file, 
and a separate program is run to carry out the canonical convolution in any desired range of the values of $p$. 

The advantage of using the NZ method is that it allows one to find the values of the observables for all values of $p$ in a single run,  averaged over many runs to get good statistical behavior.  This was particularly useful here because we did not have a good idea in advance of the behavior of the ratios $R_1$ and $R_2$ and where the curves for different $N$ crossed. 

\section{Results}

 \subsection{Disordered jammed sphere packings}
 
\begin{figure}[htbp] 
   \centering
   \includegraphics[width=3.0in]{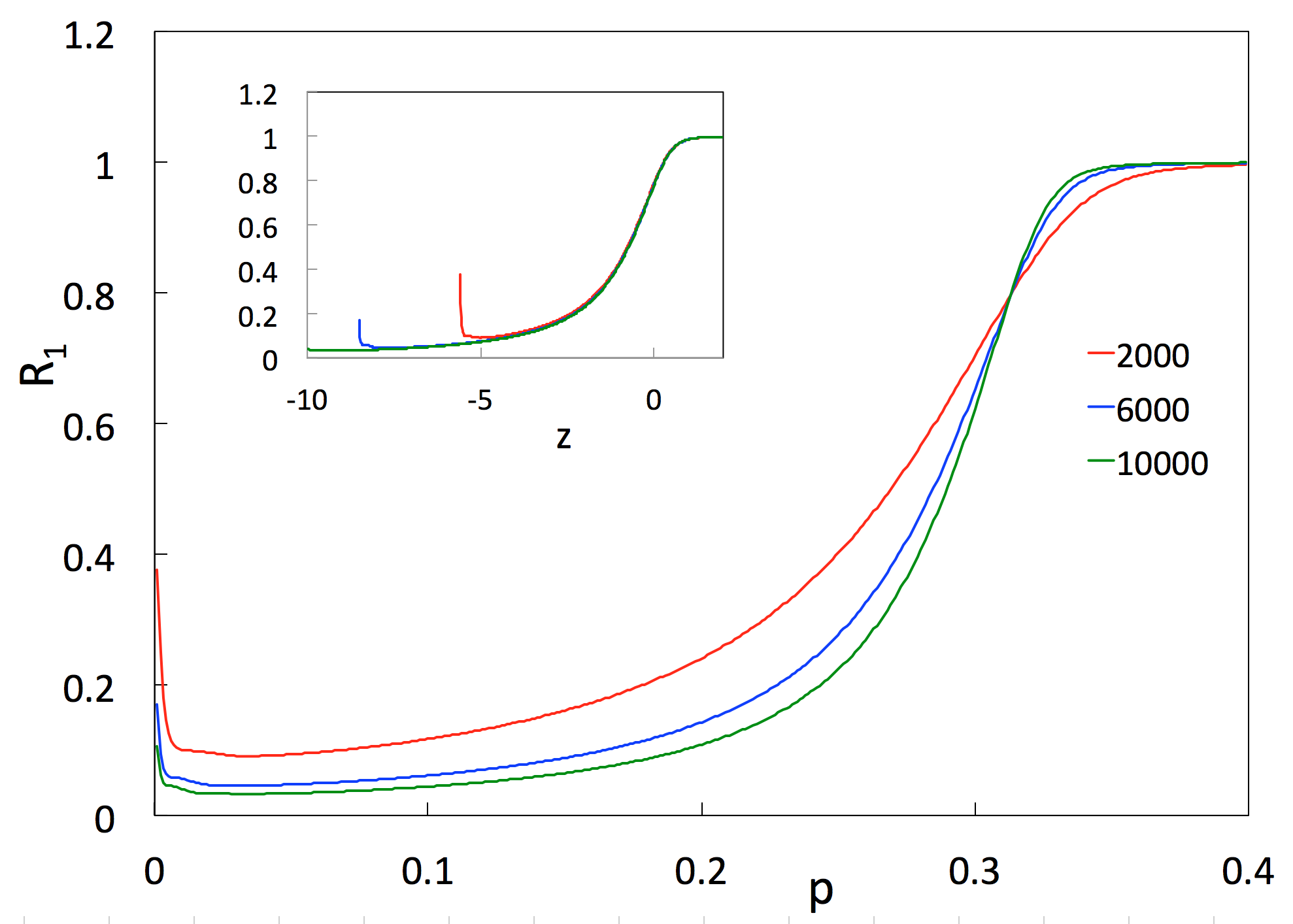} 
   \includegraphics[width=3.0in]{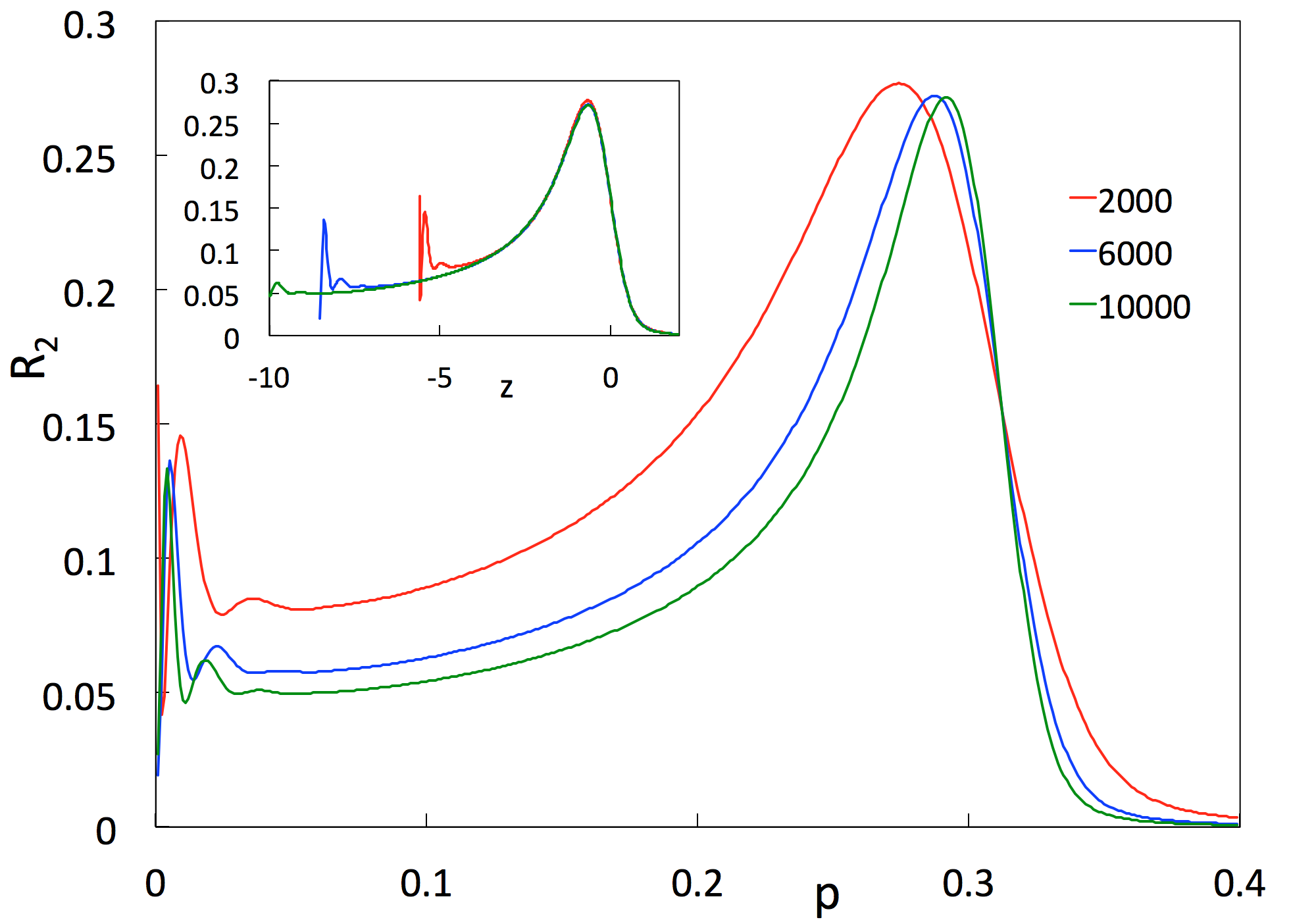} 
   \caption{Universal ratios $R_1$ (upper) and $R_2$ f(lower) vs.\ $p$ for the three samples of random packing for site percolation.  Insets show scaling plots of $R_i$ vs.\ $z = (p-p_c) L^{1/\nu}$, with $p_c = 0.3116$, $L = N^{1/3}$ and $\nu = 0.876$, showing collapse to a universal curve. } \label{fig:RsiteZT}
\end{figure}

We carried out 10,000,000 simulations on the three systems, $N = 2000, 6000$ and $10000$. 
Figure \ref{fig:RsiteZT} shows plots of the quantites $R_1$ and $R_2$ as a function of $p$ for site percolation for each $N$.  It can be seen that the curves appear to cross at a single point $p \approx 0.3116$, whereas the maxima occur below this point and at different values for different size systems.  Insets show the collapse of these curves for different size systems in a scaling plot.  Similar curves are found for bond percolation.

\begin{figure}[htbp] 
   \centering
   \includegraphics[width=3in]{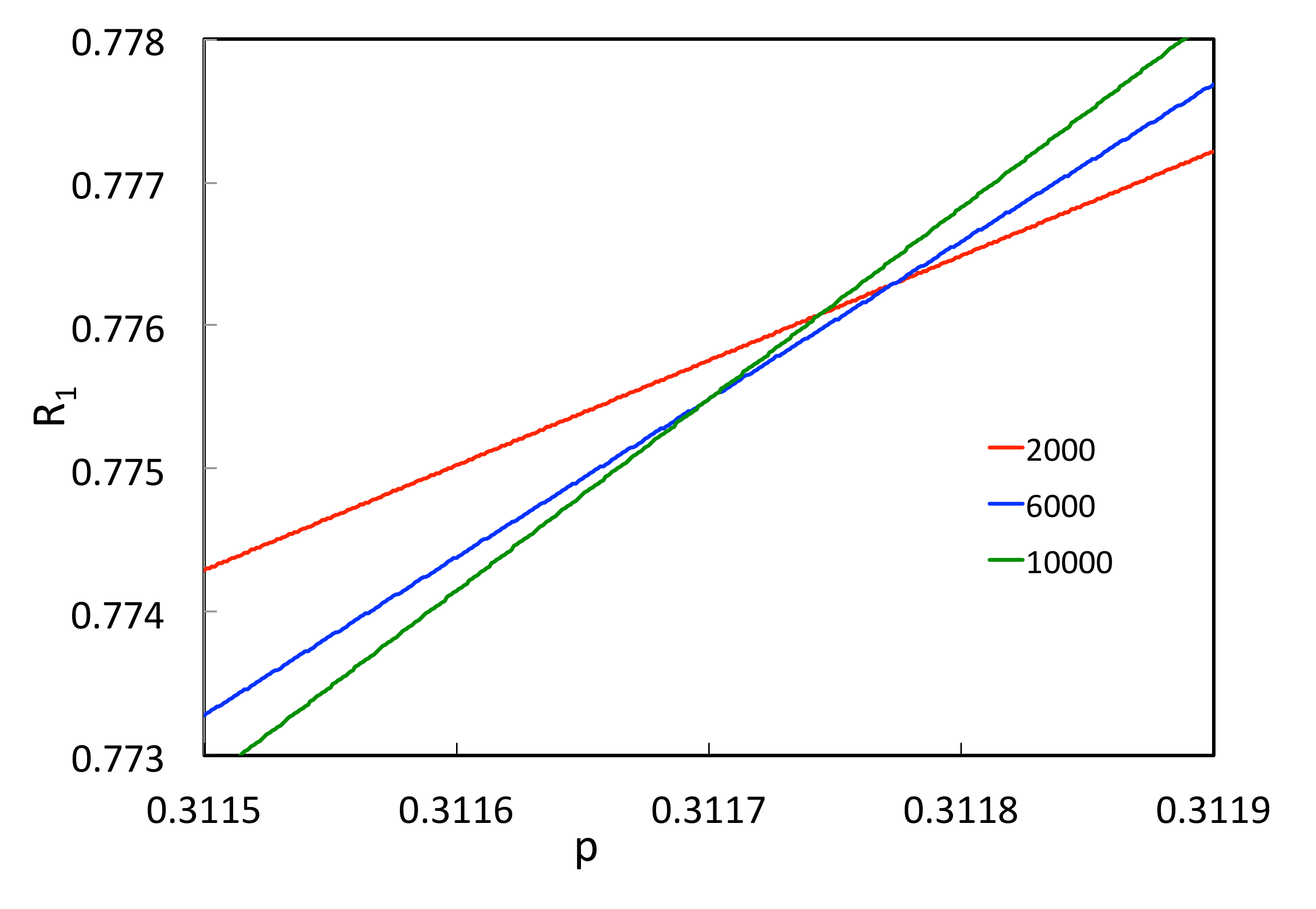} 
   \caption{Expanded view of $R_1(p)$ from Fig.\ \ref{fig:RsiteZT} for site percolation of the jammed spheres, showing that the curves do not cross precisely at a single point. }
   \label{fig:R1vsp}
\end{figure}

\begin{figure}[htbp] 
   \centering
   \includegraphics[width=3.0in]{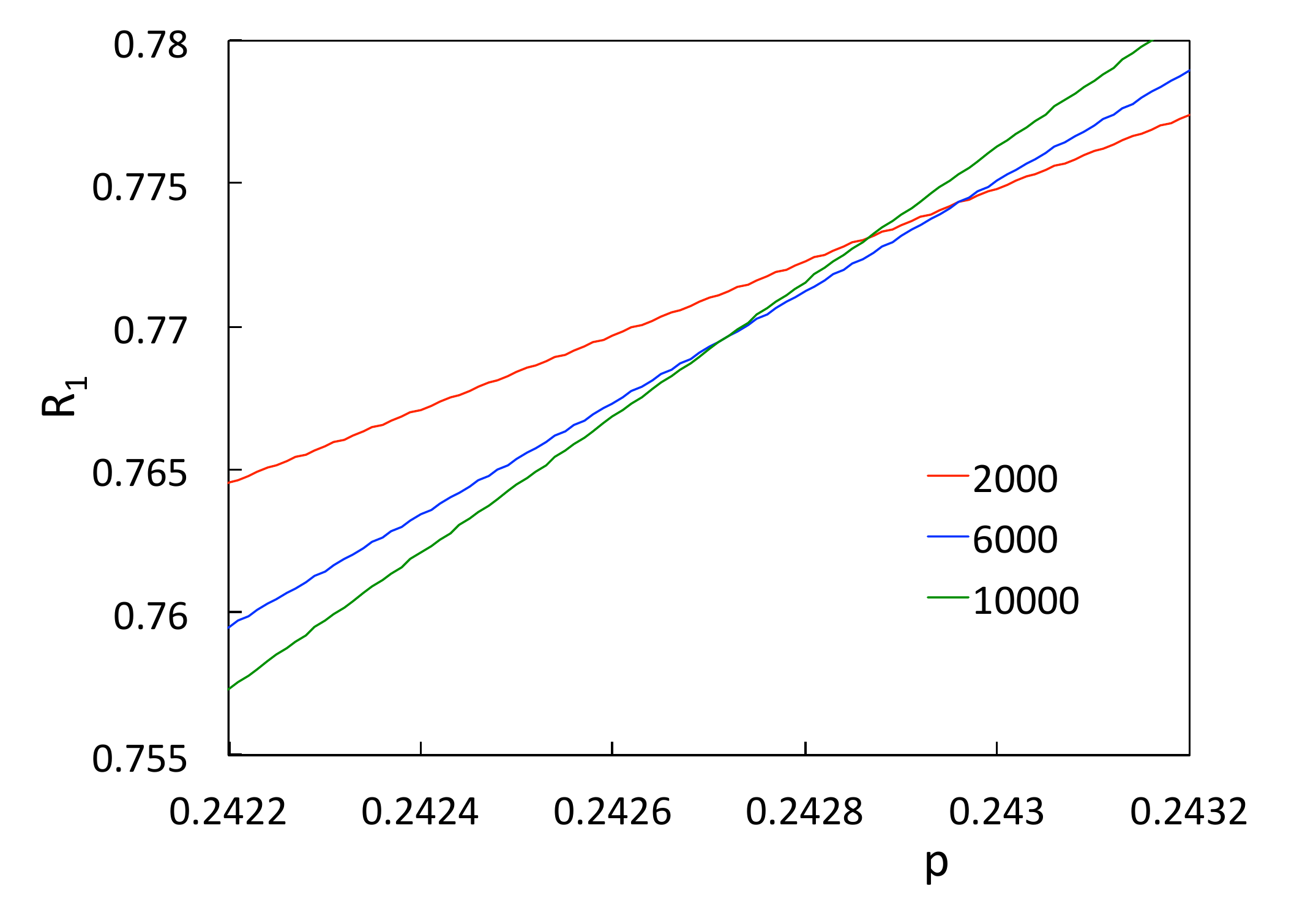} 
   \caption{Expanded plot of $R_1(p)$ for bond percolation of the disordered isostatic jammed spheres .}
   \label{fig:spheresbondR1}
\end{figure}

\begin{figure}[htbp] 
   \centering
   \includegraphics[width=3.0in]{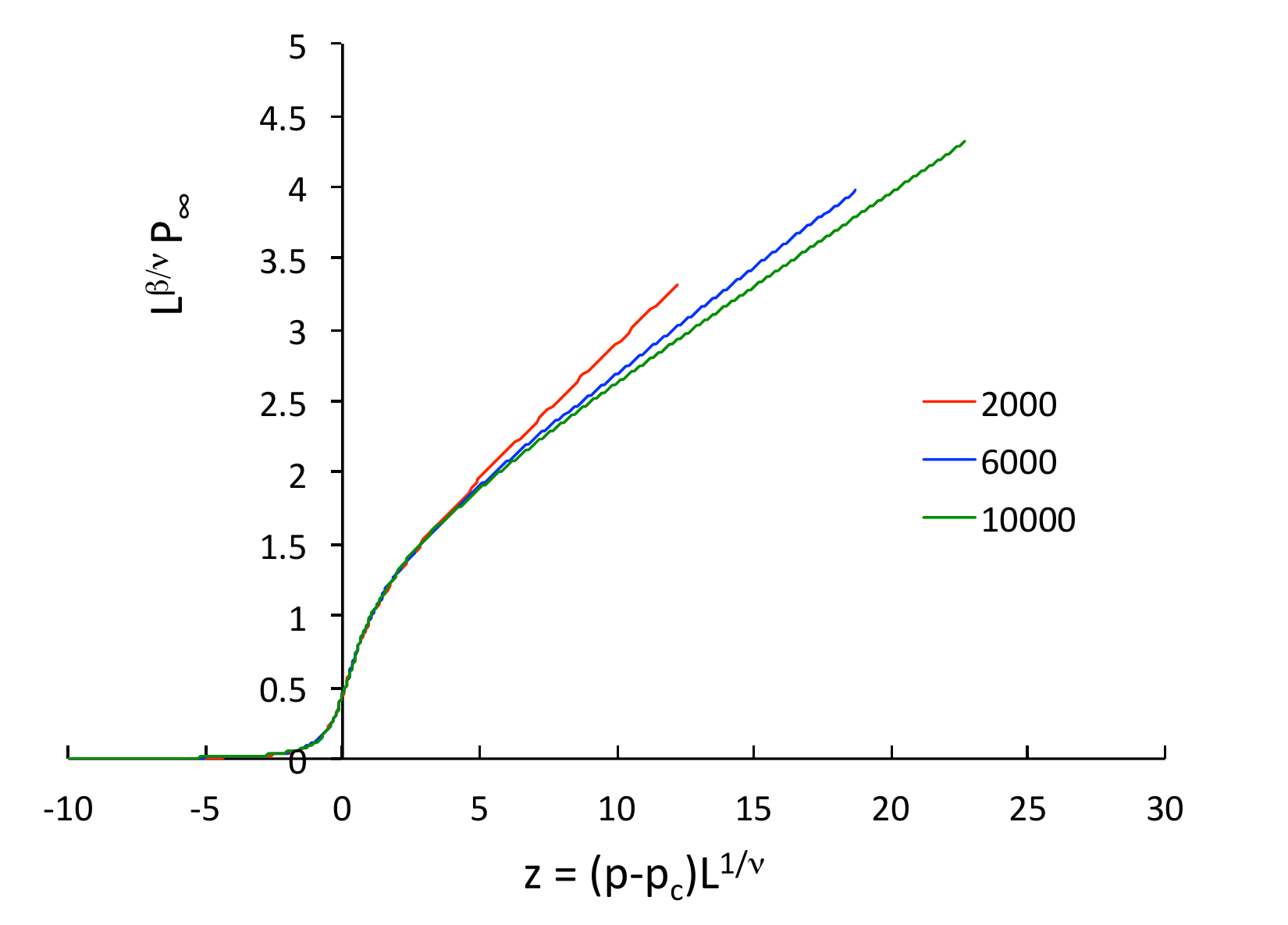} 
   \caption{Scaling plot of $L^{\beta/\nu} P_\infty$ vs.\ $(p - p_c) L^{1/\nu}$ where $P_\infty = \langle s_\mathrm{max}/N \rangle$ for site percolation on the packed spheres, with $p_c = 0.3116$ and $L = N^{1/3}$, showing collapse to a universal curve for $p$ near $p_c$.  Values of $N$ are given in the legend.}
   \label{fig:PinfnityZTsite}
\end{figure}

Figure \ref{fig:R1vsp} shows the expansion of the curves of $R_1$ in the very narrow range $0.3115 \le p \le 0.3119$, and here it can be seen that the curves do not cross exactly at a single point but at different points, due to corrections-to-scaling contributions and perhaps also statistical fluctuations.  The curves of $N=2000$ and $6000$ cross at $p = 0.31178$, and the curves of $N = 6000$ and $N = 10000$ cross at $p = 0.3117$.  We do not have enough points to make a precise extrapolation, and these data might be biased because we have only one sample of each lattice, but a reasonable extrapolation suggests  $p_c = 0.3116(3)$. 

Figure \ref{fig:spheresbondR1} shows the crossing curves for bond percolation, from which we estimate the threshold to be 0.2424(3).   

In Fig.\ \ref{fig:PinfnityZTsite} we show a scaling plot for the probability a point belongs to the largest clusters, $P_\infty = \langle s_\mathrm{max}/N \rangle$.  We assume $P_\infty$ satisfies the usual scaling relation \cite{StaufferAharony94} $P_\infty = L^{\beta/\nu} F((p-p_c) L^{1/\nu})$ where $\beta= 0.4181,$ and $\nu= 0.876$ are standard 3D scaling exponents \cite{BallesterosEtAl99,XuWangLvDeng14,HuBloteZiffDeng14} and $F(z)$ is the scaling function.  We see a good collapse of the data.

 

\subsection{Comparison to Coniglio and Stauffer (2D)}

As a check of our program and to make contact with previous results, we also considered a two-dimensional (2D) site-percolation system with open boundary conditions of size 290 x 290, exactly the system studied by Coniglio and Stauffer in 1980 \cite{ConiglioStauffer80}.   They published the values of two quantities at $p = 0.593$ (the value for the threshold assumed at that time): $\langle s_\mathrm{max} ^2 \rangle/\chi = 7.3$ and $\langle s_\mathrm{max} \rangle^2/\chi = 6.4$, where $\chi = S - \langle s_\mathrm{max} ^2 \rangle$ is the susceptibility, defined as the mean square size leaving out the largest cluster $S_\mathrm{max}$.  Their results imply $R_2 = 7.3/6.4-1 = 0.141$.   These authors generated 28,000 samples, which they believed to be the most extensive simulation of a single percolation problem at that time.   Running our program on a laptop computer for a few days, we simulated 100,000,000 samples.
With the NZ results, we could calculate these quantities at any value of $p$.  Taking $p = 0.593$, we find $\langle s_\mathrm{max} ^2 \rangle/\chi = 7.307(1)$ and  $\langle s_\mathrm{max} \rangle^2/\chi = 6.481(1)$, consistent with Coniglio and Stauffer's values, and yielding $R_2 = 7.306/6.481 - 1 = 0.1274(1)$.  Taking $p = 0.592746 = p_c$, we find  $\langle s_\mathrm{max} \rangle^2/\chi = 7.040(1)$ and  $\langle s_\mathrm{max} \rangle^2/\chi = 6.228(1)$ yielding $R_2 = 0.1304(1)$.  Notice that by changing $p$ a relatively small amount we changed the values of these quantities significantly. 

We also looked at the finite-size scaling of these quantities by considering systems of various sizes; the corrections to  $\langle s_\mathrm{max} ^2 \rangle/\chi$, $\langle s_\mathrm{max} \rangle^2/\chi$, and $R_2$  seem to all scale as $L^{-2}$, and we project for $L \to \infty$, $R_2(p_c) \to 0.1302(1)$ for 2D.  This compares to the 3D value on a periodic cube, $R_2=0.155$.


\subsection{Cubic systems}

\begin{figure}[htbp] 
   \centering
   \includegraphics[width=3in]{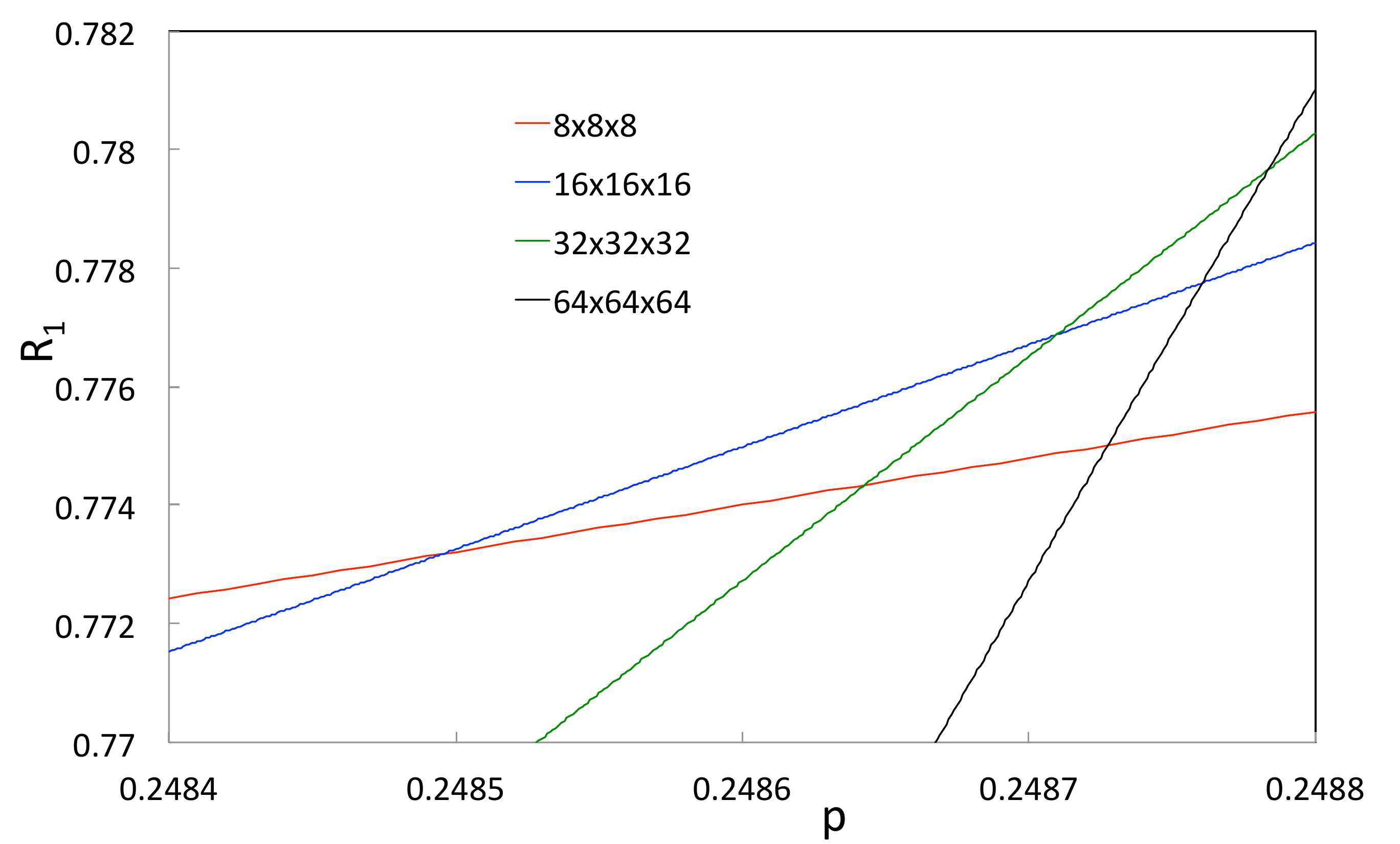} 
   \caption{Expanded view of $R_1$ vs.\ $p$ for bond percolation on a cubic lattice with dimensions given in the legend.  Here the threshold is $p_c = 0.248812$ \cite{WangZhouZhangGaroniDeng13,LorenzZiff98}. }
   \label{fig:cubicbond}
\end{figure}

For comparison, we also studied $R_1$ and $R_2$ for bond percolation on the cubic lattices, $L \times L \times L$, with $L = 8, 16, 32$ and 64, simulating, $7.5\cdot10^7$, $1.6\cdot10^8$, $1.4\cdot10^8$, and $3\cdot10^5$ samples each, respectively.  In Fig.\ \ref{fig:cubicbond} we show the crossing points for $R_1$ on an expanded scale.  They suggest a value of $R_1 \approx 0.784$ at $p_c = 0.248812$, consistent with the value if $R_1 \approx 0.78$ for the packed spheres at criticality, supporting the universality of this quantity.  We also found similar scaling behavior as seen in the packed sphere system.

\section{Conclusions}

We found that the site threshold 0.3116(3) is numerically identical (within errors) to that of the simple cubic lattice 0.311608.  This means that if one started with conducting/insulating spheres in a simple cubic arrangement at the critical site threshold 0.311608, and then jiggled them under pressure to put them into the denser disordered jammed state, then the system will remain at or very close to the critical percolation point.  This seems rather surprising, because while the average coordination number (6) does not change, the distribution of coordination numbers changes from exactly 6 to range from 4 to 12, and generally speaking, spreading out the distribution of neighbors causes the threshold to change---usually drop.
However, we would still expect these thresholds to be rather close, and it is likely that the precise agreement is a coincidence.

The simple cubic lattice has crystallographic symmetry and long-range order that is absent in MRJ. The similarity in thresholds is a statement about the insensitivity of the threshold to the precise details of the short- and long-range order provided that the average contact numbers are the same.

For bond percolation, we find $p_c = 0.2424(3)$, somewhat below the cubic system value of $p_c = 0.248812$ \cite{WangZhouZhangGaroniDeng13,LorenzZiff98}.  Evidently, the coincidence of thresholds for site percolation does not occur for bond percolation.

In Table \ref{table:thresholds} we compare the thresholds found here with other 3D systems with a coordination number of 6, both regular lattices (the kagome stack and simple cubic) and more disordered ones (the dice stack and the icosahedral Penrose lattice \cite{ZakalyukinChizhikov05}).  Having a distribution of coordination number $z$ is seen to lower the threshold, but this does not apply to the case of site percolation on the packed spheres.  Because of the higher density of the spheres, the critical volume fraction of the spheres, $\phi_c=0.199$, is the highest of this group. 

Note that the icosahedral Penrose lattice has a range of coordination numbers with an average also of 6, and has site threshold of 0.285 and a bond threshold of 0.225 \cite{ZakalyukinChizhikov05}; these thresholds were found by an approximate method and the expected errors are note clear.

The results presented here were based on measurements of three carefully constructed samples of jammed particles.  Future work could make use of more samples of each and of larger systems to find thresholds to higher precision.

For the universal ratios, we find common critical values of  $R_1 \approx 0.77$ and $R_2 \approx 0.155$ and for both the jammed spheres and cubic lattice, and for both site and bond percolation.  These are all similar periodic systems with a cubic boundary; showing commonality of these (shape-dependent) universal amplitude ratios yields a further verification that these systems are in the same universality class.

Besides studying spheres, a great deal of work has been done studying the maximum jammed state of other systems (e.g., \cite{DonevCisseSachsVarianoStillingerConnellyTorquatoChaikin04,WouterseLudingPhilipse09,JiaoStillingerTorquato10,JiaoTorquato11,HajiAkbariEngelGlotzer11,ChenJiaoTorquato14,BauleMakse14,SchallerEtAl15}). 
 These systems will also possess a jammed contact network---for example, in the tetrahedron system the mean number of nearest-neighbors is 12  \cite{ChenJiaoTorquato14}---and it would be interesting and practically useful to study their percolation thresholds as well.  Finding shapes that have low values of the critical volume fraction $\phi$ would be useful in minimizing the amount of conducting material needed in a composite for percolation to take place, for example.

 These results are also relevant for drug tablet design where percolation considerations are known to control the release of drugs \cite{SiegelKostLanger89,DamavandiZiff16}, and here there is a large interest in modeling drug release by simulation \cite{JiangLionbergerEtAl11}.    An extension of this work for applications of drug design would be to confine the spheres in a bounded system, such as a wall or a large sphere.  The behavior of the system near these boundaries would be relevant for predicting drug delivery from the surface of the tablet.  One would also be interested in the properties of the clusters of non-soluble particles that are released as a consequence of the dissolution of the soluble particles.

\acknowledgments

S. T. was supported by the National Science Foundation under Grant No.\ DMS-1211087.  R. Z. was supported by the U.S. Food and Drug Administration contract HHSF223201510157C.


\begin{table*}[h]
  \centering
  \caption{Site and bond percolation thresholds  of the jammed sphere packing and some other lattices with an average coordination number  $z$ of 6.  $f$ = filling factor and  $\phi_c=f p_c^\mathrm{site}$.  *$=$this work.}
	\begin{tabular}{ | l | l | l | l | l | l | l | }
  \hline
    Lattice & $z$& $\overline z$ & $p_c^\mathrm{site}$ & $f$ & $\phi_c$ & $p_c^\mathrm{bond}$ \\ \hline 
    kagome stack &6 &6	&0.3346(4)  \cite{vanderMarck97}& 0.453450& 0.1517&0.2563(2)  \cite{vanderMarck97}\\ 
    simple cubic &6 &6	& 0.311608 \cite{WangZhouZhangGaroniDeng13,LorenzZiff98}&0.523599&0.163158& 0.248812 \cite{WangZhouZhangGaroniDeng13,LorenzZiff98}\\  
MRJ sphere packings & 4--12 &6	&0.3116(3)*& 0.639 &0.1990*& 0.2424(3)*\\ 
 dice stack &5, 8  &6&0.2998(4)  \cite{vanderMarck97}&0.604500&0.1813&0.2378(4)  \cite{vanderMarck97}\\ 
icosahedral Penrose &4--10, 12&6	&0.285 \cite{ZakalyukinChizhikov05}&0.628630
\cite{Henley86}&0.179&0.225  \cite{ZakalyukinChizhikov05}\\ \hline
\end{tabular}
\label{table:thresholds}
\end{table*}


\bibliography{ZiffTorquato}

\begin{thebibliography}{10}

\bibitem{TorquatoStillinger10}
S.~Torquato and F.~H. Stillinger.
\newblock Jammed hard-particle packings: From {Kepler to Bernal} and beyond.
\newblock {\em Rev. Mod. Phys.}, 82:2633--2672, 2010.

\bibitem{StaufferAharony94}
Dietrich Stauffer and Ammon Aharony.
\newblock {\em Introduction to Percolation Theory, 2nd ed.}
\newblock CRC press, 1994.

\bibitem{TakeuchiInoue12}
Akira Takeuchi and Akihisa Inoue.
\newblock Critically-percolated, cluster-packed structure in {Cu--Zr} binary
  bulk metallic glass demonstrated by molecular dynamics simulations based on
  plastic crystal model.
\newblock {\em Materials Trans.}, 53(6):1113--1118, 2012.

\bibitem{FrolovPivkinaAleshin97}
Yu.~V. Frolov, A.~N. Pivkina, and A.~A. Aleshin.
\newblock Percolation phenomena in combustion of condensed systems.
\newblock {\em Chem. Phys. Reports}, 16(9):1603--1619, 1997.

\bibitem{ChuNiuLiuWang03}
Shaojun Chu, Qiang Niu, Xinyu Liu, and Xin Wang.
\newblock Site percolation in randomly packed spheres.
\newblock {\em Chinese J. Process Engineering}, 2003:413--418, 2003.

\bibitem{SiegelKostLanger89}
Ronald~A. Siegel, Joseph Kost, and Robert Langer.
\newblock Mechanistic studies of macromolecular drug release from macroporous
  polymers. {I.} {E}xperiments and preliminary theory concerning completeness
  of drug release.
\newblock {\em J. Controlled Release}, 8(3):223 -- 236, 1989.

\bibitem{LeuenbergerHolmanUsteriWinzap89}
H.~Leuenberger, L.~Holman, M.~Usteri, and S.~Winzap.
\newblock Percolation theory, fractal geometry and dosage form design.
\newblock {\em Pharm. Acta Helv.}, 64:34--39, 1989.

\bibitem{Leuenberger99}
Hans Leuenberger.
\newblock The application of percolation theory in powder technology.
\newblock {\em Advanced Powder Technology}, 10(4):323--352, 1999.

\bibitem{LeuenbergerBonnyKolb95}
H.~Leuenberger, J.~D. Bonny, and M.~Kolb.
\newblock Percolation effects in matrix-type controlled drug release systems.
\newblock {\em International Journal of Pharmaceutics}, 115(2):217--224, 1995.

\bibitem{SorianoCaraballoMillanPineroMelgozaRabasco98}
M.C. Soriano, I.~Caraballo, M.~Millán, R.T. Piñero, L.M. Melgoza, and A.M.
  Rabasco.
\newblock Influence of two different types of excipient on drug percolation
  threshold.
\newblock {\em International Journal of Pharmaceutics}, 174(1–2):63--69,
  1998.

\bibitem{MillanCaraballo06}
M\'onica Mill\'an and Isidoro Caraballo.
\newblock Effect of drug particle size in ultrasound compacted tablets:
  Continuum percolation model approach.
\newblock {\em International Journal of Pharmaceutics}, 310(1–2):168--174,
  2006.

\bibitem{KimuraPuchkovBetzLeuenberger07}
Go~Kimura, Maxim Puchkov, Gabriele Betz, and Hans Leuenberger.
\newblock Percolation theory and the role of maize starch as a disintegrant for
  a low water-soluble drug.
\newblock {\em Pharmaceutical Development and Technology}, 12(1):11--19, 2007.

\bibitem{LorenzZiff98}
Christian~D. Lorenz and Robert~M. Ziff.
\newblock Universality of the excess number of clusters and the crossing
  probability function in three-dimensional percolation.
\newblock {\em J. Phys. A: Math. Gen.}, 31(40):8147, 1998.

\bibitem{AcharyyaStauffer98}
Muktish Acharyya and Dietrich Stauffer.
\newblock Effects of boundary conditions on the critical spanning probability.
\newblock {\em Int. Jour. Mod. Phys. C}, 09(04):643--647, 1998.

\bibitem{BallesterosEtAl99}
H.~G. Ballesteros, L.~A. Fernandez, V.~Mart\'in-Mayor, A.~Munoz Sudupe,
  G.~Parisi, and J.~J. Ruiz-Lorenzo.
\newblock Scaling corrections: site percolation and ising model in three
  dimensions.
\newblock {\em J. Phys. A: Math. Gen.}, 32(1):1--13, 1999.

\bibitem{TranYooStahlheberSmall13}
Jonathan Tran, Ted Yoo, Shane Stahlheber, and Alex Small.
\newblock Percolation thresholds on three-dimensional lattices with three
  nearest neighbors.
\newblock {\em J. Stat. Mech.: Th. Exp.}, 2013(05):P05014, 2013.

\bibitem{XuWangLvDeng14}
Xiao Xu, Junfeng Wang, Jian-Ping Lv, and Youjin Deng.
\newblock Simultaneous analysis of three-dimensional percolation models.
\newblock {\em Frontiers Phys.}, 9(1):113--119, 2014.

\bibitem{WangZhouZhangGaroniDeng13}
Junfeng Wang, Zongzheng Zhou, Wei Zhang, Timothy~M. Garoni, and Youjin Deng.
\newblock Bond and site percolation in three dimensions.
\newblock {\em Phys. Rev. E}, 87:052107, 2013.

\bibitem{Malarz15}
Krzysztof Malarz.
\newblock Simple cubic random-site percolation thresholds for neighborhoods
  containing fourth-nearest neighbors.
\newblock {\em Phys. Rev. E}, 91:043301, Apr 2015.

\bibitem{JerauldScrivenDavis84}
G.~R. Jerauld, L.~E. Scriven, and H.~T. Davis.
\newblock Percolation and conduction on the 3d {V}oronoi and regular networks:
  a second case study in topological disorder.
\newblock {\em J. Phys. C: Solid State}, 17(19):3429, 1984.

\bibitem{ZakalyukinChizhikov05}
R.~M. Zakalyukin and V.~A. Chizhikov.
\newblock Calculations of the percolation thresholds of a three-dimensional
  (icosahedral) {P}enrose tiling by the cubic approximant method.
\newblock {\em Crystallography Reports}, 50(6):938--948, 2005.

\bibitem{Henley86}
C.~L. Henley.
\newblock Sphere packings and local environments in penrose tilings.
\newblock {\em Phys. Rev. B}, 34:797--816, 1986.

\bibitem{BormanGrekhovTroninTronin15}
V.~D. Borman, A.~M. Grekhov, V.~N. Tronin, and I.~V. Tronin.
\newblock Continuum percolation on nonorientable surfaces: the problem of
  permeable disks on a klein bottle.
\newblock {\em J. Phys. A: Math. Th.}, 48(47):475002, 2015.

\bibitem{Jacobsen14}
Jesper~Lykke Jacobsen.
\newblock High-precision percolation thresholds and {Potts}-model critical
  manifolds from graph polynomials.
\newblock {\em J. Phys. A: Math. Th.}, 47(13):135001, 2014.

\bibitem{Jacobsen15}
Jesper~Lykke Jacobsen.
\newblock Critical points of potts and o( n ) models from eigenvalue identities
  in periodic temperley-lieb algebras.
\newblock {\em J. Phys. A: Math. Th.}, 48(45):454003, 2015.

\bibitem{DamavandiZiff15}
Ojan~Khatib Damavandi and Robert~M Ziff.
\newblock Percolation on hypergraphs with four-edges.
\newblock {\em J. Phys. A: Math. Th.}, 48(40):405004, 2015.

\bibitem{HajiAkbariHajiAkbariZiff15}
Amir Haji-Akbari, Nasim Haji-Akbari, and Robert~M. Ziff.
\newblock Dimer covering and percolation frustration.
\newblock {\em Phys. Rev. E}, 92:032134, Sep 2015.

\bibitem{ScullardJacobsen16}
Christian~R. Scullard and Jesper~Lykke Jacobsen.
\newblock Potts-model critical manifolds revisited.
\newblock {\em Journal of Physics A: Mathematical and Theoretical},
  49(12):125003, 2016.

\bibitem{FitzpatrickMaltSpaepen74}
J.~P. Fitzpatrick, R.~B. Malt, and F.~Spaepen.
\newblock Percolation theory and the conductivity of random close packed
  mixtures of hard spheres.
\newblock {\em Phys. Lett. A}, 47(3):207 -- 208, 1974.

\bibitem{OttaviClercGiraudRoussenqGuyonMitescu78}
H.~Ottavi, J.~Clerc, G.~Giraud, J.~Roussenq, E.~Guyon, and C.~D. Mitescu.
\newblock Electrical conductivity of a mixture of conducting and insulating
  spheres: an application of some percolation concepts.
\newblock {\em J. Phys. C: Solid State}, 11(7):1311, 1978.

\bibitem{Powell79}
M.~J. Powell.
\newblock Site percolation in randomly packed spheres.
\newblock {\em Phys. Rev. B}, 20:4194--4198, 1979.

\bibitem{Powell80}
M.~J. Powell.
\newblock Site percolation in random networks.
\newblock {\em Phys. Rev. B}, 21:3725, 1980.

\bibitem{ToryEtAl73}
E.~M. Tory, B.~H. Church, M.~K. Tam, and M.~Ratner.
\newblock Simulated random packing of equal spheres.
\newblock {\em Canadian J. Chem. Eng.}, 51(4):484--493, 1973.

\bibitem{Matheson74}
A.~J. Matheson.
\newblock Computation of a random packing of hard spheres.
\newblock {\em J. Phys. C: Solid State}, 7(15):2569, 1974.

\bibitem{AhmadzadehSimpson82}
M.~Ahmadzadeh and A.~W. Simpson.
\newblock Critical concentrations for site percolation in the
  dense-random-packed hard-sphere bernal model.
\newblock {\em Phys. Rev. B}, 25:4633--4638, 1982.

\bibitem{OgerTroadecBideauDoddsPowell86}
L.~Oger, J.~P. Troadec, D.~Bideau, J.~A. Dodds, and M.~J. Powell.
\newblock Properties of disordered sphere packings {II. Electrical} properties
  of mixtures of conducting and insulating spheres of different sizes.
\newblock {\em Powder Tech.}, 46(2–3):133 -- 140, 1986.

\bibitem{Sunde96}
Svein Sunde.
\newblock {Monte Carlo} simulations of conductivity of composite electrodes for
  solid oxide fuel cells.
\newblock {\em J. Electrochem. Soc.}, 143(3):1123--1132, 1996.

\bibitem{BerteiNicolella11}
A.~Bertei and C.~Nicolella.
\newblock A comparative study and an extended theory of percolation for random
  packings of rigid spheres.
\newblock {\em Powder Tech.}, 213(1–3):100 -- 108, 2011.

\bibitem{TorquatoTruskettDebenedetti00}
S.~Torquato, T.~M. Truskett, and P.~G. Debenedetti.
\newblock Is random close packing of spheres well defined?
\newblock {\em Phys. Rev. Lett.}, 84:2064--2067, 2000.

\bibitem{TorquatoJiao10}
S.~Torquato and Y.~Jiao.
\newblock Robust algorithm to generate a diverse class of dense disordered and
  ordered sphere packings via linear programming.
\newblock {\em Phys. Rev. E}, 82:061302, 2010.

\bibitem{JiaoStillingerTorquato11}
Yang Jiao, Frank~H. Stillinger, and Salvatore Torquato.
\newblock Nonuniversality of density and disorder in jammed sphere packings.
\newblock {\em J. Appl. Phys.}, 109(1):013508, 2011.

\bibitem{AtkinsonStillingerTorquato13}
Steven Atkinson, Frank~H. Stillinger, and Salvatore Torquato.
\newblock Detailed characterization of rattlers in exactly isostatic, strictly
  jammed sphere packings.
\newblock {\em Phys. Rev. E}, 88:062208, 2013.

\bibitem{ScherZallen70}
Harvey Scher and Richard Zallen.
\newblock Critical density in percolation processes.
\newblock {\em J. Chem. Phys.}, 53(9):3759--3761, 1970.

\bibitem{TarasevichCherkasova07}
Y.~Y. Tarasevich and A.~V. Cherkasova.
\newblock Dimer percolation and jamming on simple cubic lattice.
\newblock {\em European Phys. J. B}, 60(1):97--100, 2007.

\bibitem{KriuchevskyiBulavinTarasevichLebovka14}
I.~A. Kriuchevskyi, L.~A. Bulavin, Yu. Tarasevich, and N.~I. Lebovka.
\newblock Jamming and percolation of parallel squares in single-cluster growth
  model.
\newblock {\em Condensed Matter Physics}, 17(3), 2014.

\bibitem{NewmanZiff00}
M.~E.~J. Newman and R.~M. Ziff.
\newblock Efficient {Monte Carlo} algorithm and high-precision results for
  percolation.
\newblock {\em Phys. Rev. Lett.}, 85:4104--4107, 2000.

\bibitem{NewmanZiff01}
M.~E.~J. Newman and R.~M. Ziff.
\newblock Fast {Monte Carlo} algorithm for site or bond percolation.
\newblock {\em Phys. Rev. E}, 64:016706, 2001.

\bibitem{AharonyStauffer97}
A.~Aharony and D.~Stauffer.
\newblock Test of universal finite-size scaling in two-dimensional site
  percolation.
\newblock {\em J. Phys. A: Math. Gen.}, 30(10):L301, 1997.

\bibitem{deSouzaTomeZiff11}
David~R. de~Souza, T\^ania Tom\'e, and Robert~M. Ziff.
\newblock A new scale-invariant ratio and finite-size scaling for the
  stochastic susceptible--infected--recovered model.
\newblock {\em J. Stat. Mech. Th. Exp.}, 2011(03):P03006, 2011.

\bibitem{LanglandsPichetPouliotSaintAubin92}
R.~P. Langlands, C.~Pichet, Ph. Pouliot, and Y.~Saint-Aubin.
\newblock On the universality of crossing probabilities in two-dimensional
  percolation.
\newblock {\em J. Stat. Phys.}, 67(3):553--574, 1992.

\bibitem{ConiglioStauffer80}
A.~Coniglio and D.~Stauffer.
\newblock Fluctuations of the infinite network in percolation theory.
\newblock {\em Lettere al Nuovo Cimento}, 28(1):33--38, 1980.

\bibitem{ConiglioStanleyStauffer79}
A.~Coniglio, H.~E. Stanley, and D.~Stauffer.
\newblock Fluctuations in the number of percolation clusters.
\newblock {\em J. Phys. A: Math. Gen.}, 12(12):L323, 1979.

\bibitem{Stauffer80}
D.~Stauffer.
\newblock Hausdorff dimension and fluctuations for the largest cluster at the
  two-dimensional percolation threshold.
\newblock {\em Zeitschrift f{\"u}r Physik B Condensed Matter}, 37(1):89--91,
  1980.

\bibitem{HuBloteZiffDeng14}
Hao Hu, Henk W.~J. Bl\"ote, Robert~M. Ziff, and Youjin Deng.
\newblock Short-range correlations in percolation at criticality.
\newblock {\em Phys. Rev. E}, 90:042106, 2014.

\bibitem{DonevCisseSachsVarianoStillingerConnellyTorquatoChaikin04}
Aleksandar Donev, Ibrahim Cisse, David Sachs, Evan~A. Variano, Frank~H.
  Stillinger, Robert Connelly, Salvatore Torquato, and P.~M. Chaikin.
\newblock Improving the density of jammed disordered packings using ellipsoids.
\newblock {\em Science}, 303(5660):990--993, 2004.

\bibitem{WouterseLudingPhilipse09}
A.~Wouterse, S.~Luding, and A.~P. Philipse.
\newblock On contact numbers in random rod packings.
\newblock {\em Granular Matter}, 11(3):169--177, 2009.

\bibitem{JiaoStillingerTorquato10}
Y.~Jiao, F.~H. Stillinger, and S.~Torquato.
\newblock Distinctive features arising in maximally random jammed packings of
  superballs.
\newblock {\em Phys. Rev. E}, 81:041304, Apr 2010.

\bibitem{JiaoTorquato11}
Yang Jiao and Salvatore Torquato.
\newblock Maximally random jammed packings of platonic solids: Hyperuniform
  long-range correlations and isostaticity.
\newblock {\em Phys. Rev. E}, 84:041309, 2011.

\bibitem{HajiAkbariEngelGlotzer11}
Amir Haji-Akbari, Michael Engel, and Sharon~C. Glotzer.
\newblock Phase diagram of hard tetrahedra.
\newblock {\em The Journal of Chemical Physics}, 135(19), 2011.

\bibitem{ChenJiaoTorquato14}
Duyu Chen, Yang Jiao, and Salvatore Torquato.
\newblock Equilibrium phase behavior and maximally random jammed state of
  truncated tetrahedra.
\newblock {\em J. Phys. Chem. B}, 118(28):7981--7992, 2014.

\bibitem{BauleMakse14}
Adrian Baule and Hernan~A. Makse.
\newblock Fundamental challenges in packing problems: from spherical to
  non-spherical particles.
\newblock {\em Soft Matter}, 10:4423--4429, 2014.

\bibitem{SchallerEtAl15}
Fabian~M. Schaller, Sebastian~C. Kapfer, James~E. Hilton, Paul~W. Cleary, Klaus
  Mecke, Cristiano~De Michele, Tanja Schilling, Mohammad Saadatfar, Matthias
  Schröter, Gary~W. Delaney, and Gerd~E. Schröder-Turk.
\newblock Non-universal {V}oronoi cell shapes in amorphous ellipsoid packs.
\newblock {\em EPL}, 111(2):24002, 2015.

\bibitem{DamavandiZiff16}
Ojan~Khatib Damavandi and Robert~M. Ziff.
\newblock Percolation model for drug release in nanoparticle composites.
\newblock {\em To be published}, 2016.

\bibitem{JiangLionbergerEtAl11}
Wenlei Jiang, Stephanie Kim, Xinyuan Zhang, Robert~A. Lionberger, Barbara~M.
  Davit, Dale~P. Conner, and Lawrence~X. Yu.
\newblock The role of predictive biopharmaceutical modeling and simulation in
  drug development and regulatory evaluation.
\newblock {\em International Journal of Pharmaceutics}, 418(2):151 -- 160,
  2011.

\bibitem{vanderMarck97}
S.~C. van~der Marck.
\newblock Percolation thresholds and universal formulas.
\newblock {\em Phys. Rev. E}, 55:1514--1517, 1997.

\end{thebibliography}

\end{document}